\documentstyle[epsf,12pt]{article}
\begin{document}
%\draft
\hfill\vbox{\baselineskip14pt
%            \hbox{\bf NIMS-00-xxx}
%            \hbox{NIMSL Preprint 00-xxx}
%            \hbox{\today}}
%            \hbox{January 1998}}
            \hbox{April 2005}}
\baselineskip20pt
\vskip 0.2cm 
\begin{center}
{\Large\bf Heat Capacity study of $\beta$-FeSi$_2$ single crystals }
\end{center} 
\vskip 0.2cm 
\begin{center}
\large Sher~Alam, T.~Nagai, Y.~Matsui
\end{center}
\begin{center}
{\it CSAG, AML, NIMS, Ibaraki 305-0044, Japan}
\end{center}
%\begin{center}
%{\it Physical Science Division, ETL, Tsukuba, Ibaraki 305, Japan}
%\end{center}
\vskip 0.2cm 
\begin{center} 
\large Abstract
\end{center}
\begin{center}
\begin{minipage}{14cm}
\baselineskip=18pt
\noindent
%%%%%%%%%%%%%%%%%%%%%%%%%%%%%%%%%%%%%%%%%%%%%%%%%%%%%%%%%%%%%
% This is the abstract
%\begin{abstract}
%Ins
Heat Capacity of needle-like [length=5mm, diameter=1 mm]
$\beta$-FeSi$_{2}$ single crystal, grown by chemical vapor 
transport has been measured. Two anomalies are found, a
broad deviation centered around 160 K and a clear deviation 
at a temperature of 255 K approximately. We have attempted to 
relate these to the anomalies previously reported in the case 
of the resistivity data.  The Transient 
Thermoelectric Effect [TTE] results lead us to the inference
that the system under goes from single carrier system to at
least two carrier system at 220 K-our heat capacity results
seem to provide further independent evidence for this transition
in this system.  
%\end{abstract}
\end{minipage}
\end{center}
\vfill
\baselineskip=20pt
\normalsize
\newpage
\setcounter{page}{2}
%------------------------------------------------------------------------ 
% Section:
%\section{Introduction}
%\twocolumn
% Update January 2003	
\section{Introduction}
	Among the silicides $\beta$-FeSi$_{2}$, a semiconductor,
is an interesting and promising material for several reasons.
Broadly speaking $\beta$-FeSi$_{2}$ is environmentally friendly,
and it is also compatible with existing silicon technology.
From the technological point of view the photoelectric properties
of this and related materials may be utilized in optoelectronic
devices that can be integrated into silicon technology. 
Silicon dominates the microelectronics industry. However
silicon is a poor emitter of light due to its indirect
band-gap, this leads to efficiencies of 0.01-0.1 $\%$
even for complex silicon based LED structures. Thus there
are many efforts to remedy this situation. In this regard
semiconducting silicides\cite{bor99} offer several advantages.
In particular, Iron disilicide $\beta$-FeSi$_{2}$ is a
promising material for optoelectronic applications plus it is
as already mentioned it is  environmentally friendly or
Kankyo semiconductor. Specifically it emits light at 
1.55 $\mu$m [0.8 eV] which is the value required for SiO$_{2}$
optical fiber communications.  
However there are some hurdles in fabricating  
$\beta$-FeSi$_{2}$ films on Si-substrates using 
Molecular Beam Epitaxy. The diffusion of iron into Si
substrate is perhaps the most difficult to overcome.
In short $\beta$-FeSi$_{2}$ is attractive as potential
constituents in optical and thermoelectric devices.
But, a main problem remains that the semiconducting and
physical properties are very sensitive to sample preparation
\cite{beh2001,udo2001}. Yet another issue is that band calculations
suggest a strong coupling between band edge states to the
lattice \cite{chr1990}, implying low carrier mobility. In contrast, 
magneto-transport experiments \cite{aru1994} indicate the existence of
high mobility carriers in addition to the ones with low
mobility. Recently Hara et al.\cite{hara2003} reported
a ''phase-transition'' around 220 K based on resistivity
data. 

In this paper we report on our experimental measurements
of the heat capacity of needle-like single crystal of
$\beta$-FeSi$_{2}$, grown by chemical vapor transport 
method \cite{osa2001}. We find an anomaly at 255 K in the 
heat capacity data, which provides support to resistivity 
transport data. There also
seems to be another ''anomaly'' at lower temperature of 100 K.
From the Transient Thermoelectric Effect [TTE] results we can infer
that the system under goes from single carrier system to at
least two carrier system at 220 K \cite{hara2003}. 

\section{Experimental}
\begin{figure}
\epsfbox{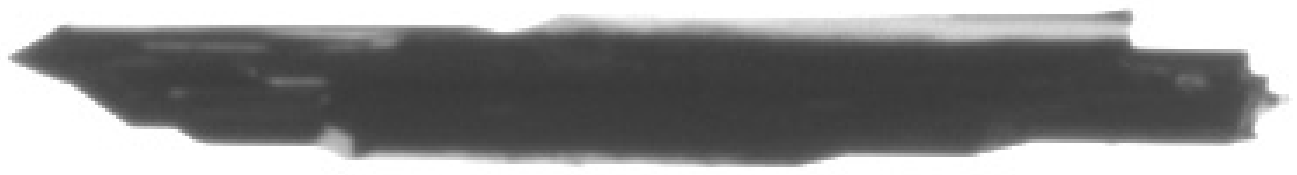}
\caption{Photograph of the needle-like single crystals 
of $\beta$-FeSi$_{2}$, with approximate dimensions, length=5mm,
diameter=1 mm.}
\label{fig1}
\end{figure}

Needle-like single crystals of $\beta$-FeSi$_{2}$ were grown by chemical
vapor transport [CVT] method \cite{osa2001}. The sample preparation is made
from the gas phase, using iodine as a carrier gas, single phase needle-like
$\beta$-FeSi$_{2}$ bulk crystals are obtained, with typical dimensions
5-10 mm length and 1mm diameter. The $\beta$-FeSi$_{2}$ needle-like crystal
used in heat capacity experiment is shown in Fig.~\ref{fig1}. The approximate 
dimensions of the crystal are 5mm length and 1mm diameter. The  
measured value of the mass was 1.26 $\pm 0.01$ mg. 

The Heat Capacity [HC] was measured using
Quantum Design Physical Property Measurement System [PPMS].
 The HC of the sample is calculated by subtracting
the addenda measurement from the total heat capacity measurement. The
total HC is the measurement of the HC of the sample, the grease,
and the sample platform. The two measurements-one with and one without
the sample on the sample platform are necessary for accuracy. In order
to ensure the further accuracy of our results, we conducted the
experiment several times. We note that automatic subtraction of the 
addenda, at each sample temperature measurement is performed. 
We have performed the measurement in the temperature range 295-38 K.
%%%%%%%%%%%%%%%%%%%%%%%%%%%%%%%%%%%%%%%%%%%%%%%%%%%%%%%
\section{Results and Discussion}
\begin{figure}
\epsfbox{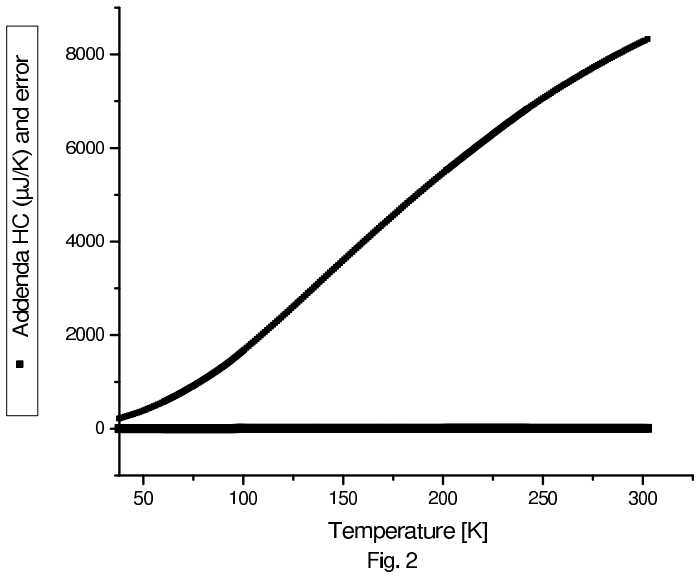}
\caption{The Addenda HC [$\mu$ J/K] and the error used for the sample.}
\label{fig2}
\end{figure}

\begin{figure}
\epsfbox{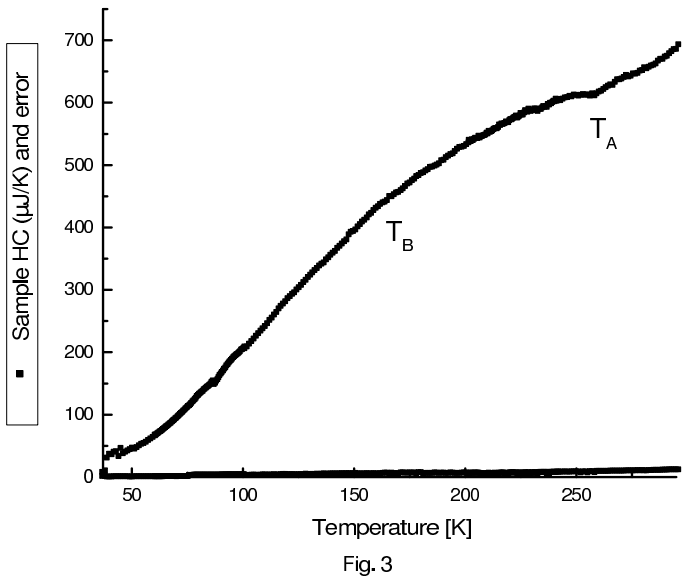}
\caption{The sample HC [$\mu$ J/K] and the corresponding 
error.}                     
\label{fig3}
\end{figure}

\begin{figure}
\epsfbox{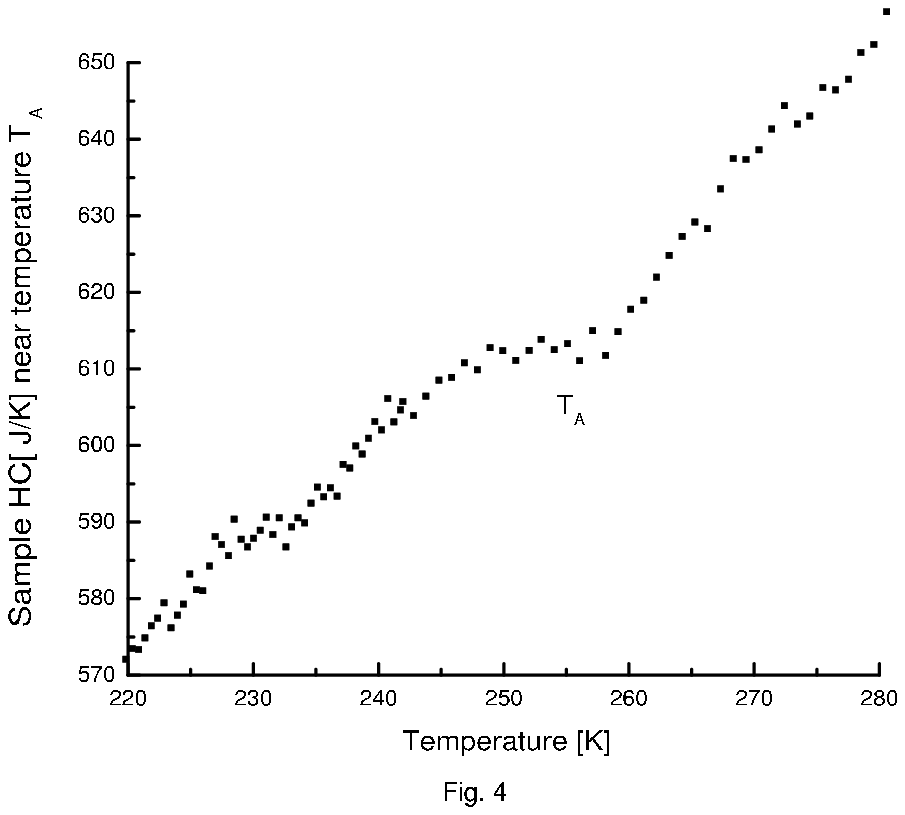}
\caption{The main HC anomaly in more detail.}
\label{fig4}
\end{figure}

\begin{figure}
\epsfbox{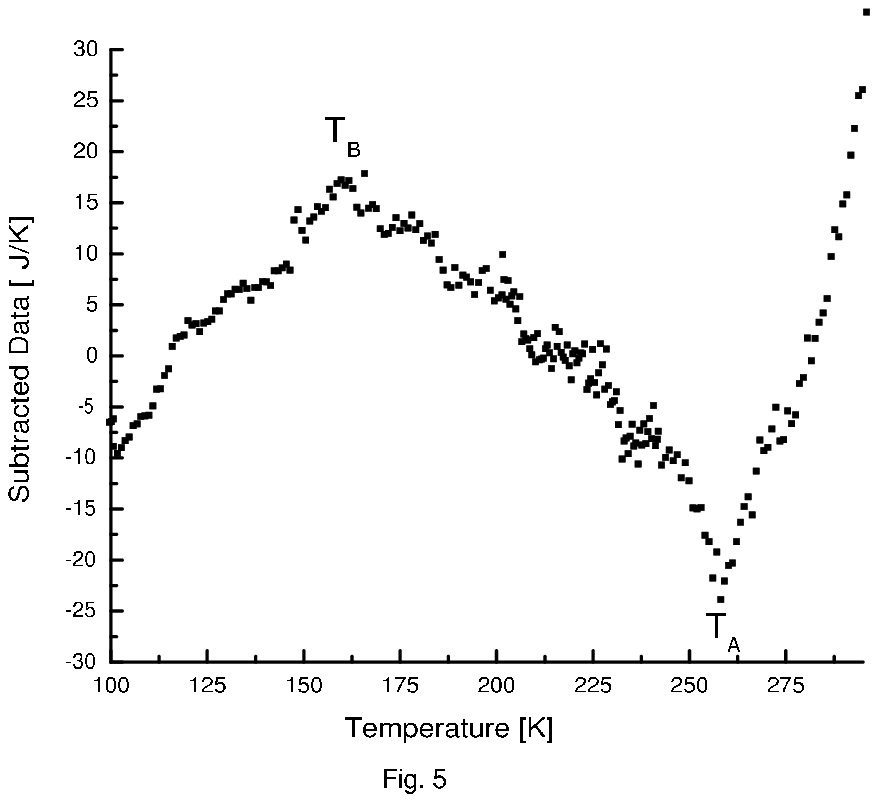}
\caption{The subtracted data,$\Delta C$ [$\mu$ J/K] showing 
the anomalies at T$_{_A}$  and T$_{_{B}}$ for the sample.}
\label{fig5}
\end{figure}

Fig.~\ref{fig1} shows the $\beta$-FeSi$_{2}$ needle-like crystal
used in heat capacity experiment. The crystal is approximately 5 mm in
length and with a 1 mm diameter. The mass is 1.26$\pm 0.01$ mg.
Fig.~\ref{fig2} through Fig.~\ref{fig5} show the results of our 
HC measurements. The the results of the addenda measurement, along with
the corresponding error are shown in Fig.~\ref{fig2}. It is clear that the
addenda data is smooth as it should be, without any jumps
which could confuse the interpretation of the sample data.
As can be clearly seen the error is negligible.
The sample HC and the corresponding error are shown in Fig.~\ref{fig3}.
The error is very small compared to the value of the HC. The
two anomalies are indicated by temperatures T$_{_A}$ and
T$_{_B}$. To clarify the behaviour of these anomalies, we
show in detail the anomaly at temperature T$_{_A}$ in Fig.~\ref{fig4},
and show the subtracted HC $\Delta C$ in Fig.~\ref{fig5}.
The approximate deviation at the temperature T$_{_A}$ is on the
order of 6.6-8.3 $\%$. The broad deviation centered around
T$_{_B}$ has a maximum deviation of 6.25 $\%$ peak to peak,
with respect to the value of HC at T$_{_B}$.

                  What is the origin of the anomaly at
T$_{_A}$? The resistivity data clearly shows an anomaly at
a temperature of 220 K \cite{hara2003}, there is an inflection in
the $\rho$-T curve, which is indicative of ''phase-transition''
rather than a cross-over. Now our HC {\it measurements strongly 
suggest the existence of a ''phase-transition''}. Incidentally
this clarifies the discrepancy between band-structure calculations
and magneto-transport experiments mentioned previously in the 
introduction-since one can argue that band structure changes,
with some bands having very sharp edges, as a result of the
phase transition. In addition the TTE measurements also 
confirm this scenario-these measurements indicate that that
the system undergoes from single carrier system to at least
two carrier system at approximately 220 K. The anomaly which
is centered around approximately 160 K, Fig.~\ref{fig5}, lies
in region 100 K $\le$ T $\le$ 250 K. This roughly coincides
with Region II ($\approx$ 90 K $\le$ T $\le$ 220 K) in the 
notation of Hara et al.\cite{hara2003}. In this region
the behaviour of resistivity $\rho$ indicates that the
system behaves as a degenerate semiconductor with density
of states, which is finite even at absolute zero.
In addition from the Hall measurements, in this region
non-linear Hall effect was observed, which is suggestive
that the system behaves as a multiple carrier one. This also
fits nicely with the slow broad deviation, shown here in
Fig.~\ref{fig5} as the  change of HC from the smooth 
background- covering the whole region  between 100 K and 250 K.
%%%%%%%%%%%%%%%%%%%%%%%%%%%%%%%%%%%%%%%%%%%%%%%%%%%%%%%%%%%
\section{Conclusions}
In conclusion we have measured the HC of needle-like single
crystals of $\beta$-FeSi$_{2}$ single crystal, grown by chemical 
vapor transport. We found two anomalies, which can be related
to the ones found previously in the measurements of resistivity
and TTE. Thus providing evidence and clarification of the
resistivity and TTE data. Very importantly if one assumes
that the ''phase transition'' at T$_{A}$ changes the 
band-structure in a significant way, so that some bands
develop sharp edges, then this can clarify the discrepancy
between the band-structure calculations and magneto-transport
measurements. Our finding is significant also in the sense
that it provides an independent confirmation of the previous
results using an entirely different technique. 
%%%%%%%%%%%%%%%%%%%%%%%%%%%%%%%%%%%%%%%%%%%%
%\section{}
%\subsection{Raw Data and Standard}
%\subsection{Main Edge}
%\subsection{Pre-edge Region}
%\subsection{Difference Spectra}
%\section{Discussion}
%\section{Conclusions}
%\end{itemize} 
%%%%%%%%%%%%%%%%%%%%%%%%%%%%%%%%%%%%%%%%%%%%%%%%%%%%%%%%%%%%
%\begin{enumerate}
%\item{}
%\end{enumerate} 
%%%%%%%%%%%%%%%%%%%%%%%%%%%%%%%%%%%%%%%%%%%%%%%%
%In a metallic system close to a metal-insulator transition
%the electronic properties play an important and a crucial role.
%%%%%%%%%%%%%%%%%%%%%%%%%%%%%%%%%%%%%%%%%%%
%\newpage
\section*{Acknowledgements}
The sample was obtained from Masato Osamura, and Sher Alam thanks him
for that. Sher Alam also thank Y. Makita for introducing him
to $\beta$-FeSi$_{2}$. We will also like to thank Dr. E.Takayama 
Muromachi and his group at AML, NIMS for the use of their PPMS system. 
The work of Sher Alam is supported by Dr. Matsui electron
microscopy group. The HC experiments were conducted in April 2005.

%%%%%%%%%%%%%%%%%%%Ins%%%%%%%%%%%%%Figures%%%%%%%%%%%%%%
%\section*{Figure Captions}
%\begin{center}
%Figure Captions
%\end{center}
\newpage
%%%%%%%%%%%%%%%%%%%%%%%%%%%%%%%%%%%%%%%%%%%%%%%%%%%%%%%%%%
%%%%%%%%%%%%%%%%%%%%%%%%%%%%%%%%%%%%%%%%%%%%%%%%%%%%%%%%%%%
%%%%%%%%%%%%%%%%%%%%%%%%%%%%%%%%%%%%%%%%%%%%%%%%%%%%%%%%%%
%%%%%%%%%%%%%%%%%%%%%%%%%%%%%%%%%%%%%%%%%%%%%%%%%%%%%%
%%%%%%%%%%%%%%%%%%%%%%%%%%%%%%%%%%%%%%%%%%%%%%%%%%%
%\includegraphics{IAfig1.pdf}
%\newpage
%\epsfbox{IAfig1.eps}

%\epsfbox{IAfig2f.eps}

%\epsfbox{IAfig3.eps}

%\epsfbox{IAfig4.eps}

%\epsfbox{IAfig5.eps}
\end{document}